\documentclass[a4paper,preprint]{emulateapj}
\usepackage{amstext}
\usepackage{amsmath}

\begin{document}

\title{Radio Flares and the Magnetic Field Structure in GRB Outflows}

\author{Jonathan Granot\altaffilmark{1} and
Gregory B. Taylor\altaffilmark{1,}\altaffilmark{2}}

\altaffiltext{1}{Kavli Institute for Particle Astrophysics and
Cosmology, Stanford University, P.O. Box 20450, MS 29, Stanford, CA
94309; granot@slac.stanford.edu} \altaffiltext{2}{National Radio
Astronomy Observatory, P.O. Box O, Socorro, NM 87801;
gtaylor@aoc.nrao.edu}

\begin{abstract}

The magnetic field structure in gamma-ray burst (GRB) outflows is of
great interest as it can provide valuable clues that might help pin
down the mechanism responsible for the acceleration and collimation
of GRB jets. The most promising way of probing this magnetic field
structure is through polarization measurements of the synchrotron
emission from the GRB ejecta, which includes the prompt $\gamma$-ray
emission and the emission from the reverse shock. Measuring
polarization in $\gamma$-rays with current instruments is extremely
difficult: so far there is only one claim of detection (a very high
degree of linear polarization in GRB 021206) which despite the
favorable conditions remains highly controversial. The emission from
the reverse shock that propagates into the ejecta as it is
decelerated by the ambient medium peaks in the optical on a time
scale of tens of seconds (the so called `optical flash') and
dominates the optical emission up to about ten minutes after the
GRB. Unfortunately, no polarization measurements of this optical
emission have been made to date. However, after the reverse shock
finishes crossing the shell of GRB ejecta, the shocked ejecta cools
adiabatically and radiates at lower and lower frequencies. This
emission peaks in the radio after about one day, and is called the
`radio flare'. We use VLA data of radio flares from GRBs to
constrain the polarization of this emission. We find only upper
limits for both linear and circular polarization. Our best limits
are for GRB 991216, for which we find $3\;\sigma$ upper limits on
the linear and circular polarization of $7\%$ and $9\%$,
respectively. These limits provide interesting constraints on
existing GRB models.  Specifically, our results are hard to
reconcile with a predominantly ordered toroidal magnetic field in
the GRB outflow together with a `structured' jet, where the energy
per solid angle drops as the inverse square of the angle from the
jet axis, that is expected in models where the outflow is Poynting
flux dominated.

\end{abstract}


\section{Introduction}
\label{int}

The detection of linear polarization at the level of $\sim 1\%-3\%$
in the optical afterglow emission of several gamma-ray bursts
\citep[GRBs; see][for a review]{Covino03} has been widely considered
as a confirmation that synchrotron emission is the dominant
radiation mechanism, at least in the afterglow stage. Synchrotron
radiation is also believed to be the dominant emission mechanism in
the prompt $\gamma$-ray emission and in the emission from the
reverse shock, although the observational support for this is not as
strong as for the afterglow. Soon after the first detection of
linear polarization in the afterglow emission
\citep{Covino99,Wijers99} it has been realized that the temporal
evolution of the polarization (both the degree of polarization $P$
and its position angle $\theta_p$) can probe the magnetic field
structure in the emitting region, as well as the structure and the
dynamics of GRB jets \citep{GL99,Sari99}.

While the polarization properties of the afterglow emission have
received relatively wide attention
\citep{LP98,GW99,GL99,Sari99,ML99,IN01,GK03,Rossi04}, the polarization
of the prompt GRB emission received very little
attention\footnote{\citet{GK03} had, however, pointed out that an
  ordered magnetic field in the ejecta could produce a high degree of
  polarization, $P\lesssim 60\%$, in the emission from the GRB ejecta,
  which includes the prompt $\gamma$-ray emission, and the emission
  from the reverse shock (the `optical flash' and `radio flare').}
before the detection of a very high degree of linear polarization
($P=80\%\pm 20\%$) in the prompt $\gamma$-ray emission of GRB 021206
\citep{CB03}. Although this detection is highly controversial
\citep{RF04,Wigger04}, it has dramatically raised the interest in
the polarization properties of the prompt emission. Two main
explanations have been suggested for the production of tens of
percent of polarization in the prompt $\gamma$-ray emission: (i)
synchrotron emission from an ordered magnetic field in the ejecta
\citep{GK03,CB03,LPB03,Granot03}, and (ii) a viewing angle just
outside the sharp edge of a jet, $\theta_0<\theta_{\rm
  obs}\lesssim\theta_0+1/\Gamma_0$ \citep{Gruzinov99,Waxman03}, where
$\theta_0$ and $\Gamma_0$ are the initial half-opening angle and
Lorentz factor of the jet, respectively. The second explanation can
work with either synchrotron emission
\citep{Waxman03,Granot03,NPW03} or inverse-Compton scattering of
external photons \citep{SD95,EL03,Lazzati04a}.  In all the above
cases, the polarization can be a good fraction of the maximal
polarization for synchrotron emission, $P_{\rm max}\approx
60\%-70\%$.\footnote{For inverse Compton scattering of external
  photons, the local polarization from a given point on the image can
  approach $100\%$, however the averaging over the unresolved image
  reduces the observed polarization to values only slightly
  higher than those for synchrotron emission.}

The second explanation requires a narrow jet,
$\theta_0\Gamma_0\lesssim{\rm a\ few}$, in order to have a reasonable
probability of viewing the jet at an appropriate angle,
$\theta_0<\theta_{\rm obs}\lesssim\theta_0+1/\Gamma_0$. Since a larger
value of $\theta_0\Gamma_0$ is inferred for most GRBs
\citep{PK01a,PK01b}, this would imply little or no polarization in
most cases, where our viewing angle is inside the jet, $\theta_{\rm
  obs}<\theta_0$. The first explanation, however, would imply a high
polarization in all GRBs, if indeed they have a magnetic field that
is ordered over angular scales $\gtrsim 1/\Gamma_0$.

The best way to reliably measure the polarization of the emission
from the GRB ejecta is by using the emission from the reverse shock,
which includes the optical flash and the radio flare. While no
polarization measurements have been made so far in the optical
flash, the polarization of the radio flare can be inferred from
existing data. In \S \ref{RS} we discuss the existing candidates for
radio flare emission in GRBs and the evidence that this emission is
indeed from the reverse shock. In \S \ref{obs} we use archival data
to derive upper limits on the linear and circular polarization of
the radio flare emission, and in \S \ref{prop_ext} we show that
external propagation effects in the host galaxy or circumburst
environment are not likely to cause a significant depolarization the
radio emission. The upper limits on the polarization are contrasted
with the predictions of different theoretical models in \S
\ref{theory}. In \S \ref{prop_int} we discuss propagation effects
inside the source. We discuss the magnetic field configuration in
the GRB ejecta in light of our results in \S \ref{dis} and give our
conclusions in \S \ref{conc}.

\section{Is the Radio Emission from the Reverse Shock?}
\label{RS}

In order to draw conclusions regarding the magnetic field structure
in the GRB ejecta from the polarization measurements of the radio
flares, it is important to have some confidence that the radio
emission is indeed from the original ejecta, whose electrons were
heated by the reverse shock and then cooled adiabatically
\citep{SP99}. The best candidates are GRBs 990123
\citep{Kulkarni99a}, 991216 \citep{Frail00} and 020405
\citep{Berger03}. A radio flare was also reported for GRB 970828
\citep{Djorgovski01}, however in this case the source was detected
in the radio at only one epoch and at one frequency, and there is no
good evidence that suggests that this emission arises from the
reverse shock. Also, the signal to noise in this case is rather bad,
and does not give meaningful constraints on the polarization.
Therefore, we shall concentrate on the other three GRBs for which
there is better evidence that the radio flare emission is indeed
from the reverse shock, and for which the quality of the data allows
us to place interesting limits on the polarization.

In GRB 990123 the early radio emission (up to a day or two) has been
found to agree very well with the expectations for the reverse shock
emission \citep{SP99,Kulkarni99a,NP04}, especially when also taking
into account the prompt optical emission from this GRB
\citep{Akerlof99} which is also successfully explained as emission
from the reverse shock. Furthermore, it is hard to explain the early
radio emission as arising from the forward shock. Therefore, in GRB
990123 we have good reason to believe that the $8.46\;$GHz radio
measurement at $t=1.25\;$days that is used in \S \ref{obs} to derive
limits on the polarization was indeed dominated by emission from the
reverse shock.

In GRB 991216 the radio emission in the first few days is
inconsistent with the forward shock emission that is responsible for
the optical and X-ray emission \citep{Frail00}, and a different
emission component is required. \citet{Frail00} suggested either
reverse shock emission, which naturally explains the early radio
emission, or a two component jet model, where the early radio
emission is from a spherical component with $E_{\rm iso}\sim
10^{54}\;$erg running into a very low external density, $n\sim
10^{-4}\;{\rm cm^{-3}}$. Such a low density is inconsistent with
broad band afterglow fits to GRB 991216 \citep{PK01a,PK01b} which
give a density higher by $\sim 4-5$ orders of magnitude. Also, the
energy in the spherical component is very high, $\sim
10^{53}-10^{54}\;$erg, which is $\sim 2-3$ orders of magnitude
higher than the values inferred for other GRBs
\citep{Frail01,PK01b}. Thus, the reverse shock is probably the best
candidate for the early radio emission in GRB 991216, although it is
hard to conclusively rule out other explanations.

In GRB 020405 the early rapid decay and spectral slope in the radio,
$F_\nu\propto t^{-1.2\pm0.4}\nu^{-0.3\pm0.3}$, provide a reasonably
good case that the early radio emission (within the first few days)
is dominated by the reverse shock \citep{Berger03}.

\section{Limits on the Polarization from Radio Flares}
\label{obs}

We use archival VLA observations in order to measure the
polarization of the radio flare emission discovered from GRBs to
date.  The VLA observations are typically short observations
($\sim$30 minutes) that were squeezed into the VLA schedule for
these targets of opportunity. As such, there is insufficient
calibration information from these short runs alone to allow for
good polarization calibration in the standard manner using a
calibrator observed over a wide range in parallactic angle.  To
overcome this difficulty we combined the VLA observations of the
afterglows with calibration data taken from 1--4 days before or
after the run.  The instrumental leakage terms are stable on
time-scales of months, so we are confident that an optimal
calibration has been obtained.  As a check, the polarization
calibration was applied to the absolute flux calibrator observed on
the same day as the radio flare (3C286 or 3C138), and the results
were found to be consistent with their well-known properties.

We obtain only upper limits on the linear and circular polarization
of three radio flares.  Our results are summarized in Table 1. It is
worth noting that so far only upper limits have been found for the
radio polarization from GRB afterglows \citep{Taylor04, Taylor05}.
In the best case of the bright GRB 030329, the $3\;\sigma$ limits
are $<$1\% \citep{Taylor04}.

\newcommand{\rb}[1]{\raisebox{1.5ex}[0pt]{#1}}
\begin{deluxetable}{|l|ccccc|}
\tabletypesize{\footnotesize} \tablecaption{Limits on the
Polarization of Radio Flares in GRBs} \tablewidth{0pt} \tablehead{
\colhead{GRB} & \colhead{$t$ (days)} & \colhead{$t_j$ (days)} &
\colhead{$\Pi_L$} & \colhead{$\Pi_C$} & \colhead{$F_\nu$ ($\mu$Jy)}
}
\startdata
990123 & 1.25 & $\approx 2$ & $<23\%$ & $<32\%$ & $242\pm 26$
\\ \hline
 & 1.49 &  & $<11\%$ & $<17\%$ & $946\pm 56$
\\
991216 & 2.68 & $\sim 2$ & $<9\%$ & $<15\%$ & $634\pm 26$
\\
 & 1.49, 2.68 &  & $<7\%$ & $<9\%$ &  $715\pm 25$
\\ \hline
020405 & 1.19 & $\sim 1-2$ & $<11\%$ & $<19\%$ & $492\pm 29$
\enddata
\tablecomments{The $3\;\sigma$ upper limits on the linear
polarization, $\Pi_L=(Q^2+U^2)^{1/2}/I$, and circular polarization,
$\Pi_C=V/I$, of radio flares from GRBs, where $Q$, $U$, $V$ and $I$
are the Stokes parameters. In the third line for GRB 991216 we
combine the two epochs from the first two lines in order to obtain a
better limit on the polarization. These limits were derived using
VLA observations at $8.46\;$GHz.}

\end{deluxetable}

\section{Propagation Effects External to the Source}
\label{prop_ext}

Propagation effects might reduce the intrinsic linear polarization
below detectable levels.  A Faraday screen produced by ionized gas
and magnetic fields can cause gradients in the observed polarization
angle across the source, leading to depolarization if the intrinsic
source size or the resolution element of the telescope is large
compared to the gradients.  The rotation measures (RM) can be
related to the line-of-sight magnetic field, $B_{\|}$, by
\begin{equation}
{\rm RM} = 812\int\limits_0^L \left(\frac{n_{\rm e}}{1\;{\rm
cm^{-3}}}\right)\left(\frac{B_{\|}}{\rm mG}\right)
\left(\frac{dl}{\rm 1\;pc}\right)~{\rm rad\;m}^{-2}\ ,
\end{equation}
where the upper limit of integration, $L$, is the distance from the
emitting source to the end of the path through the Faraday screen
along the line of sight. To produce a 90$^\circ$ rotation at
$8.4\;$GHz requires a RM of $1200\;{\rm rad\;m^{-2}}$.

In our Galaxy the RM can reach up to $\sim 1000\;{\rm rad\;m^{-2}}$
at very low Galactic latitudes, while at high Galactic latitudes it
drops to a few$\,\times\, 10\;{\rm rad\;m^{-2}}$ \citep{SNKB81}. At
low Galactic latitudes \citet{Clegg92} find a RM difference of up to
$180\;{\rm rad\;m^{-2}}$ on scales of $1^\circ$ which translates to
a linear scale of $\sim 100-200\;$pc and gradient of $\Delta{\rm
RM}/\Delta L_\perp\sim 1\;{\rm rad\;m^{-2}\;pc^{-1}}$ in the RM.
Since in our case the source size is $\sim (1-3)\times 10^{-2}\;$pc,
this would correspond to a negligible rotation measure gradient
across the image, $\Delta{\rm RM}\sim (1-3)\times 10^{-2}\;{\rm
rad\;m^{-2}}$, some 5 orders of magnitude too low to cause
significant depolarization. The host galaxies of GRBs are not
thought to be greatly different from our own Galaxy \citep{Bloom02}.

Molecular clouds in our Galaxy typically exhibit variation of
 $\Delta{\rm RM}\sim 18-30\;\;{\rm rad\;m^{-2}}$ in
the RM over scales of $\sim 2\;$pc \citep{WR04}. This corresponds to
$\Delta{\rm RM}/\Delta L_\perp\sim 9-15\;{\rm rad\;m^{-2}\;pc^{-1}}$
and $\Delta{\rm RM}\sim 0.1-0.4\;{\rm rad\;m^{-2}}$ across the
image, which is still some 4 orders of magnitude too low to cause
significant depolarization.
Individual H{\sc ii} regions can produce enhanced RMs with
dispersions of order $\sim 50\;{\rm rad\;m^{-2}}$ on scales of $\sim
0.2\;$pc \citep{Gaensler01}, corresponding to $\Delta{\rm RM}/\Delta
L_\perp\sim 250\;{\rm rad\;m^{-2}\;pc^{-1}}$ and $\Delta{\rm RM}\sim
2.5-7.5\;{\rm rad\;m^{-2}}$ across the image, which is still some
2-3 orders of magnitude too low.

Now we consider propagation effects in the immediate environment of
the GRB, which was shaped by its progenitor. If the latter is a
massive star, the immediate environment is the pre-explosion stellar
wind which for a constant mass loss rate and wind velocity would
give a density profile $n_{\rm ext}\propto R^{-2}$, at radii $R$
that are smaller than that of the wind termination shock. Such a
density profile would imply a small deceleration radius $R_{\rm
dec}$ and a very high density at that radius, which would in turn
imply fast cooling of the reverse shock electrons and no detectable
radio flare emission at $t\sim 1\;$day. Keeping this in mind, we
shall assume a uniform external medium.

For GRB~020405, \cite{Berger03} model the optical, X-ray and radio
light curves with a uniform medium of density $n_{\rm ext}\sim
0.05\;{\rm cm^{-3}}$. Afterglow models applied to GRB~020405 predict
an intrinsic diameter of the radio emitting region at 1.2 days after
the burst of $\sim 10^{17}\;$cm. Values of the external density
inferred from broad band afterglow modeling can be as high as
$n\lesssim 30\;{\rm cm^{-3}}$ \citep{PK01a,PK01b,Yost03}. For GRB
990123 a very low density of $n_{\rm ext}\sim 10^{-3}\;{\rm
cm^{-3}}$ is inferred, while for GRB 991216 the inferred density is
more typical, $n_{\rm ext}\sim 5\;{\rm cm^{-3}}$.

Magnetic fields within the circumburst medium are not expected to
cause a large depolarization, unless there is a clump with a high
density contrast along our line of sight.
A clump much smaller than the image size would have a small effect.
A clump larger or comparable to the image size, $L_{\rm im}$, would
produce a change $\Delta{\rm RM}$ in the rotation measure across the
image which is roughly independent of the clump size, $L$, for a
given density contrast $C$. This can be seen as follows. We have
$n=Cn_{\rm ext}\propto C$ and assuming flux freezing and isotropic
compression, $B\propto n^{2/3}\propto C^{2/3}$ so that ${\rm
RM}\propto nBL\propto C^{5/3}L$, and the change in the RM across the
image is roughly $\Delta{\rm RM}\sim {\rm RM}\times L_{\rm
im}/L\propto C^{5/3}$, where the dependence on $L$ cancels out.
Furthermore, the gradient in the rotation measure across the image
would be comparable to the total rotation measure of a clump with a
size similar to that of the image, $\Delta{\rm RM}\sim{\rm
RM(L=L_{\rm im})}$, causing a relative rotation of the polarization
position angle of
\begin{equation}\label{clump}
\frac{\Delta\theta_p}{90^\circ}\lesssim 0.45\left(\frac{n_{\rm
ext}}{1\;{\rm cm^{-3}}}\right)\left(\frac{B_{\rm ext}}{10\;\mu{\rm
G}}\right)\left(\frac{2R_\perp}{10^{17}\;{\rm
cm}}\right)\left(\frac{C}{100}\right)^{5/3}\ ,
\end{equation}
where $n_{\rm ext}$ and $B_{\rm ext}$ are the number density and
magnetic field of the un-clumped external medium, and $R_\perp$ is
the apparent radius of the image on the plane of the sky. Thus a
clump with a density contrast $C\gtrsim 10^2$ and a size $\gtrsim
10^{17}\;$cm is required in order to induce significant
depolarization. We also note that if such a clump lies within a
distance of $\lesssim 1\;$pc from the progenitor star it would
produce a detectable bump in the afterglow light curve, which was
not observed for GRBs 990123, 991216 or 020405. Thus, such a clump
along our line of sight seems unlikely. It is much less likely that
such a clump would be along our line of sight in all the three GRBs
that we analyzed.

Instead of a single clump, one might envision many small clumps with
a covering factor of $f\sim 1$ (in order to cover most of the
image). The typical clump size would have to be
$L_{17}=L/(10^{17}\;{\rm cm})< 1$ in order to avoid producing
detectable bumps in the afterglow light curve. Since even for
$C_2=C/100\sim 1$ we need $L_{17}\gtrsim 1$ in order for a single
clump to produce significant depolarization (see Eq. \ref{clump}),
many small clumps are required along a random line of sight. If
their rotation of the position angle would always be in the same
direction, the required number of clumps would be $N_0\gtrsim
C_2^{-5/3}L_{17}^{-1}$. However, since the direction of rotation of
the position angle is expected to be random between different
clumps, the total rotation would on average be zero with a mean
r.m.s value of $N_0^{1/2}$ times that of a single clump. Thus a
strong depolarization would require $\gtrsim N_0^2$ clumps along a
random line of sight. Thus $N_{\rm cl}(R)L^2/R^2\gtrsim N_0^2\sim
C_2^{-10/3}L_{17}^{-2}$ where $N_{\rm cl}(R)$ is the total number of
clumps at radii $\sim R$. Thus the total mass in the clumps would be
$\sim 0.1M_\odot(n_{\rm ext}/1\;{\rm
cm^{-3}})C_2^{-7/3}L_{17}^{-1}R_{18}^2$ while there volume filling
factor would be $\sim 0.1 C_2^{-10/3}L_{17}^{-1}R_{18}^{-1}$. This
means that the clumps would hold $\sim 10
C_2^{-7/3}L_{17}^{-1}R_{18}^{-1}$ times more mass than that of the
material between the clumps. Such an extreme clumping seems highly
unlikely.

\section{The Resulting Constraints on Theoretical Models}
\label{theory}

The radio flare emission typically peaks on a time scale of $\sim
1\;$day after the GRB. This is usually similar to the jet break
time, $t_j$, in the afterglow light curve. At this time the Lorentz
factor is typically $\Gamma\sim 10$, which is much smaller than its
initial value, $\Gamma_0\gtrsim 100$. Thus a good part of the jet
(or all the jet for $t>t_j$) is visible, and the observed
polarization is an effective average value over this observed
region. The optical flash emission peaks at the deceleration time
$t_{\rm dec}$ when the Lorentz factor is close to its initial value
(unless the reverse shock is highly relativistic), and should thus
have a polarization close to that of the prompt $\gamma$-ray
emission. It could therefore provide information that is
complimentary to that from the radio flare.

The jet break time, $t_j$, plays an important role in the
polarization light curve for most theoretical models. Therefore, it
is important to know its value when comparing between theory and
observations. In GRB 990123 there is a jet break in the optical
light curve at $t_j\approx 2\;$days \citep{Kulkarni99b,PK01a}. A
similar jet break time, $t_j\approx 2\;$days, was found in the
optical light curve of GRB 991216 \citep{Halpern00,PK01a} with a
rather large uncertainty on this value. In GRB 020405 the jet break
is not readily apparent in the data.\footnote{For example,
\citet{Masetti03} find no evidence for a jet break in the light
curve up to $\sim 10\;$days. Since the most severe constraints from
our polarization measurements are for an ordered toroidal magnetic
field in the ejecta, in which case a larger value for the jet break
time ($t_j\gtrsim 10\:$days) would imply even stricter constraints
on the model, we adopt a conservative approach and use the lower
values of $t_j$ that were inferred by \citet{Berger03} and
\citet{Price03}.} \citet{Berger03} deduce $t_j\approx 0.95\;$days
from the fit to the broad band afterglow light curve, while
\citet{Price03} infer $t_j=1.67\pm0.52\;$days. Thus we use a value
of $t_j\sim 1-2\;$days for GRB 020405. In all cases the polarization
limits are at times either similar to or slightly before the jet
break time, $t\lesssim t_j$.

If the observed radio frequency is below the self absorption
frequency, $\nu<\nu_{\rm sa}$, this can significantly reduce the
polarization, for any magnetic field configuration. \citet{NP04} found
that the peak of the radio flare emission is typically due to the
passage of the self absorption frequency across the observed band.
This would imply $\nu<\nu_{\rm sa}$ and a suppression of the
polarization during the rise to the peak, and $\nu>\nu_{\rm sa}$ (no
suppression of the polarization) during the decay after the peak. This
can reduce the polarization before or around the time of the peak in
the radio flare emission. In GRBs 991216 and 020405 the radio flux is
decaying at the time of the measurement of the radio flare, and direct
measurements of the spectral slope also support an optically thin
spectrum ($\nu>\nu_{\rm sa}$), so that no suppression of the
polarization is expected due to self absorption. For GRB 990123 the
point used to measure the polarization seems to be near the peak of
the radio flare and there is no good measurement of the spectral slope
near this time.  Thus the polarization might be somewhat suppressed if
$\nu\lesssim\nu_{\rm sa}$, but probably not significantly suppressed
since the relatively high flux and the proximity to the peak of the
radio flare suggest that $\nu$ is not much smaller than $\nu_{\rm sa}$
and the two are at most comparable.

\subsection{The Ejecta Dynamics after the Reverse Shock}
\label{dyn}

Since the radio flare emission typically peaks at $t\lesssim t_j$,
the possible lateral spreading of the jet at $t\gtrsim t_j$ can be
neglected, and the jet dynamics can be reasonably approximated as
being part of a spherical flow.  After the passage of the reverse
shock, the shocked external medium approaches the \citet[][hereafter
BM]{BM76} self similar solution.  However, the shocked ejecta has a
significantly higher density than that given by the BM solution, and
thus its rest mass density becomes comparable to its internal energy
density (i.e. it becomes ``cold") much earlier on. For a mildly
relativistic reverse shock this happens soon after the reverse shock
crosses the shell, while for a relativistic reverse shock the
shocked shell is first reasonably described by the BM solution,
while it is still ``hot", but deviates from that solution once it
becomes ``cold" \citep{KS00}.

For the BM solution the value of the self similar variable $\chi$
for a fixed fluid element, that is appropriate for the original
ejecta, evolves with radius $R$ as $\chi=(R/R_0)^{4-k}$ for an
external density $\rho_{\rm ext}\propto r^{-k}$ \citep{GS02}, where
$R_0$ is the radius where it crossed the shock, which for the
original shell of ejecta is given by the deceleration radius,
$R_0\sim R_{\rm dec}$. Also, $\gamma=\Gamma\chi^{-1/2}$ where
$\Gamma\propto R^{-(3-k)/2}$ is the Lorentz factor of the fluid just
behind the shock, so that $\gamma\propto R^{-(7-2k)/2}$ for the
original ejecta. More generally one can assume some power law
dependence of the Lorentz factor on radius, $\gamma\propto R^{-g}$,
where the power law index $g$ can deviate from its value for the BM
solution, $g=7/2-k$. This implies $t\propto
R/\gamma^2\propto\gamma^{-(2g+1)/g}\propto $,
$\gamma\propto t^{-g/(2g+1)}$ and $R\propto t^{1/(2g+1)}$.

Once the shell becomes ``cold" (i.e. its rest energy exceeds its
internal energy) its dynamics would deviate from the BM solution
\citep{KS00}, however, the power law index $g$ is still bounded by
the value just behind the forward shock, $g=(3-k)/2$ (since the the
ejecta shell is lagging behind the shocked external medium) and by
the value for the BM solution, $g=7/2-k$ (since there is a larger
inertia that resists the deceleration which is driven by a similar
pressure). Thus, $(3-k)/2<g<7/2-k$ or $3/2<g<7/2$ for $k=0$ (a
uniform density external medium).\footnote{\label{SC}A stellar wind
environment is not so relevant for the radio flare since in this
case the reverse shock electrons are are fast cooling (i.e. cool
significantly due to radiative losses within the dynamical time) and
therefore the observed flux rapidly decays after the passage of the
reverse shock, and no detectable radio flare is expected
\citep{CL00,KZ03}.}

\begin{figure}
\plotone{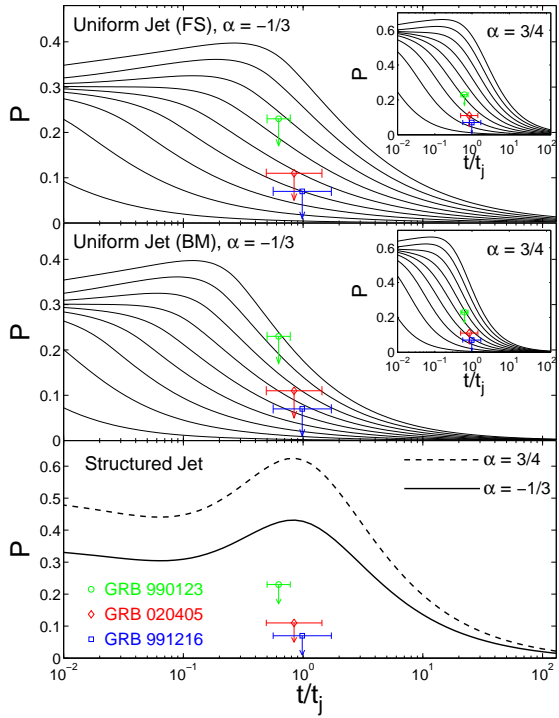}
\caption{\label{fig1}Our $3\;\sigma$ upper limits for the linear
polarization of the radio flare emission overlaid on the theoretical
polarization light curves for a toroidal magnetic field in the GRB
ejecta. The error bars represent the uncertainty in the
determination of the jet break time $t_j$ from the optical afterglow
light curve. The top two panels are for a uniform jet, and are
calculated in Appendix \ref{app1}. The different lines, from top to
bottom are for $\theta_{\rm obs}/\theta_0=0.9,\,0.8,\,...,\,0.1$.
The main figure is for $\alpha=-d\log F_\nu/d\log\nu=-1/3$ and
$P_{\rm max}=(\alpha+1)/(\alpha+5/3)=1/2$, while the inset is for
$\alpha=3/4$ and $P_{\rm max}=21/29\approx 0.724$. In the top panel
the Lorentz factor of the ejecta is assumed to remain equal to that
of the freshly shocked fluid just behind the forward shock (FS),
while the middle panel it is assumed to follow the \citet{BM76} (BM)
self similar solution. The bottom panel is for a `structured' jet
where the energy per solid angle drops as $\theta^{-2}$ outside of
some small core angle \citep[taken from][]{Lazzati04b}. In this case
$P(t/t_j)$ is practically independent of $\theta_{\rm obs}$.}
\end{figure}

At any given observed time $t>t_{\rm dec}$ the observed emission from
the reverse shock is from a somewhat smaller radius and a slightly
smaller Lorentz factor compared to the emission from the forward
shock. This can be seen as follows. Let quantities normalized to their
values at $t_{\rm dec}$ be denoted by a twiddle. From the equality of
the observed time we have $\tilde{R}_{\rm RS}=\tilde{R}_{\rm
  FS}^{(4-k)/(2g+1)}=\tilde{R}_{\rm FS}^{(2g_{\rm FS}+1)/(2g_{\rm
    RS}+1)}$ where $g_{\rm FS}=(3-k)/2$ for the forward shock and
$g=g_{\rm RS}$ for the reverse shock. Since $R\propto t^{1/(2g+1)}$
this implies that the ratio of the Lorentz factors for a fixed
observed time is $\gamma/\Gamma=\tilde{R}_{\rm FS}^{-(g_{\rm
    RS}-g_{\rm FS})/(2g_{\rm RS}+1)}=\tilde{t}^{-(g_{\rm RS}-g_{\rm
    FS})/[(2g_{\rm RS}+1)(2g_{\rm
    FS}+1)]}=\tilde{t}^{-[g-(3-k)/2]/[(4-k)(2g+1)]}$. For $k=0$ this
gives $\gamma/\Gamma=\tilde{t}^{-(g-3/2)/[4(2g+1)]}$ where the
exponent ranges between $-1/16$ and $0$ for the possible range of
$3/2<g<7/2$, which implies $\gamma/\Gamma\approx 1$.\footnote{Since
typically
  $t_{\rm dec}\sim 10-100\;$s and the radio flare emission is observed
  at $t\sim 10^5\;$s, this implies $1<\Gamma/\gamma\lesssim 1.5-1.8$.}
The somewhat smaller radius of the ejecta shell and the fact that it
is typically cold by the time of the radio flare which is typically
at $t\lesssim t_j$ suggest that a possible lateral spreading of the
jet is probably not important for the radio flare emission, and thus
it shall be neglected when calculating the polarization.

\subsection{The Implications for Different Theoretical Models}
\label{imp}

If a high polarization in the prompt GRB is caused by a viewing
angle $\theta_0<\theta_{\rm obs}\lesssim\theta_0+1/\Gamma_0$, then
the polarization of the optical flash around its peak would be
similar to that of the prompt $\gamma$-ray emission. This is since
at this time ($t\sim t_{\rm dec}$) $\Gamma\sim\Gamma_0$ and
$\theta_j\approx\theta_0$ (i.e. very little lateral expansion could
have taken place before the deceleration time). However, at the time
of the radio flare ($t\sim 1\;$day) $\Gamma\ll\Gamma_0$ and even a
modest lateral expansion would give $\Gamma_0(\theta_j-\theta_0)\gg
1$, so that the jet would occupy our line of sight. This could
significantly reduce the polarization for a magnetic field that is
random within the plane of the shock, as is expected to be produced
in relativistic collisionless shocks by the two stream instability
\citep{ML99}. Furthermore, for a shock produced magnetic field the
polarization of the radio flare should be similar to that of the
optical emission from the forward shock, which is observed to be
$P\lesssim 3\%$.\footnote{In order to detect the afterglow emission
and measure its polarization one must first detect the prompt
$\gamma$-ray emission, which requires $\theta_{\rm
obs}\lesssim\theta_0+1/\Gamma_0$. Since typically
$\Gamma_0\theta_0\gg 1$, most lines of sight would be inside the
jet, $\theta_{\rm obs}<\theta_0$.}

Such a low polarization suggests that the magnetic field is not
random in 2D, fully within the plane the shock, but is instead more
isotropic, and has a comparable component in the direction normal to
the shock \citep{GK03}. Constraints similar to our best case, GRB
991216, for a larger number of GRB, could suggest a similar
conclusion for the magnetic field in the ejecta.\footnote{A line of
sight sufficiently close to the jet axis could produce a very low
polarization. However, such viewing angles correspond to a small
solid angle, so that the probability for such lines of sight in all
of the GRBs in a reasonably sized sample is very small.}

Our upper limits on the polarization of the radio flare emission can
put much tighter constraints on models where there is an ordered
magnetic field in the GRB ejecta. If there is a magnetic field in
the ejecta that is ordered over patches of angular scale
$\theta_B\gtrsim 1/\Gamma_0$ then typically $P\sim P_{\rm max}$
during the prompt $\gamma$-ray emission and near the peak of the
optical flash, while during the radio flare the polarization can be
reduced due to averaging over $N\sim(\gamma\theta_B)^{-2}$
incoherent patches, $P\sim P_{\rm max}\times\min[1,\gamma\theta_B]$
\citep{GK03}. This implies $\theta_B\lesssim P_{\rm lim}/\gamma
P_{\rm max}$, where $P_{\rm lim}$ is our upper limit on the linear
polarization. During the radio flare $\gamma\sim 10$ so that for GRB
991216 we have $P_{\rm lim}=0.07$ and
$\theta_B\lesssim 1.4\times 10^{-2}(\gamma/10)^{-1}(P_{\rm
max}/0.5)^{-1}\;$rad. In particular, if the magnetic field is
roughly uniform over the whole jet ($\theta_B\sim\theta_0$) then
this would imply $P\sim P_{\rm max}$ \citep{GK03}, i.e. tens of
percent of polarization, which is definitely inconsistent with our
upper limits.

For a toroidal magnetic field in the ejecta, we show our upper
limits superimposed on the theoretical polarization light curves in
Fig \ref{fig1}, both for a uniform jet and for a `structured' jet.
The latter is expected in models where the GRB outflow is Poynting
flux dominated \citep{LPB03,LB04}. The polarization light curve for
a structured jet is taken from \citet{Lazzati04b}. The polarization
light curves for a uniform jet are derived in Appendix \ref{app1}.
For a uniform jet we consider the two limiting cases for the
evolution of the Lorentz factor of the ejecta after the deceleration
time (see \S \ref{dyn} for details): (i) the same as the forward
shock Lorentz factor, and (ii) following the \citet{BM76} self
similar solution. In case (ii) the Lorentz factor of the ejecta
$\gamma$ decreases slightly faster with the observed time $t$
($\propto t^{-7/16}$ instead of $t^{-3/8}$). This slightly
`contracts' the polarization light curve along the time axis. Also,
while $\gamma\theta_j(t_j)=1$ in case (i), we have
$\gamma\theta_j(t_j)=(t_j/t_{\rm dec})^{-1/16}\sim 0.65$ where
typically $t_j/t_{\rm dec}\sim 10^3$, which shifts the polarization
light curves to earlier times by a factor of $(t_j/t_{\rm
dec})^{1/7}\sim 2.7$. This can be readily seen by comparing the
upper and middle panels in Fig. \ref{fig1}.

Since $P_{\rm max}=(\alpha+1)/(\alpha+5/3)$ \citep{Granot03}, a
higher values of $\alpha$ produces a higher degree of polarization.
Nevertheless, even their lowest values for optically thin
synchrotron emission, $\alpha=-1/3$ and $P_{\rm max}=1/2$, still
produce a fairly high degree of polarization.

For a uniform jet, $P(t\sim t_j)$ significantly increases with
$\theta_{\rm obs}/\theta_0$, and goes to zero at $\theta_{\rm
obs}=0$. Thus, our upper limits on the polarization put upper limits
on $\theta_{\rm obs}/\theta_0$. These limits also depend on the
dynamical model for the GRB ejecta. For GRB 991216 we obtain
$\theta_{\rm obs}/\theta_0\lesssim 0.4$ and $0.55$ for cases (i) and
(ii), respectively.

The model that is most severely constrained by our upper limits on
the polarization is a toroidal magnetic field together with a
structured jet (see lower panel of Fig. \ref{fig1}). In this case
all of our upper limits are significantly below the predictions of
this model (by a factor of $\gtrsim 6$ for GRB 991216). Thus a
predominantly toroidal magnetic field in the GRB ejecta together
with a structured jet is hard to reconcile with our upper limits.

\section{Propagation Effects at the Source}
\label{prop_int}

The effects of propagation in plasma inside the source on the
synchrotron emission in GRBs has been recently considered
\citep[][hereafter SLW04]{MI03,SWL04}. In the early emission while
the reverse shock is still going on, these effects can suppress
linear polarization and produce up to tens of percent of circular
polarization below the self absorption frequency $\nu_{\rm sa}$, for
a magnetic field that is ordered on large scales (SLW04). We have
obtained upper limits both on the linear polarization ($\Pi_L$) and
on the circular polarization ($\Pi_C$) for the radio flare emission.
The best limits are for GRB 991216: $\Pi_L<7\%$ and $\Pi_C<9\%$
($3\;\sigma$). Our upper limits directly constrain such plasma
propagation effects.

There are some arguments which suggest that plasma propagation
effect should be relatively small and sub-dominant in the radio
flare emission. First, the suppression of linear polarization and
prevalence of circular polarization is much larger when the reverse
shock electrons are fast cooling, i.e. most electrons cool
significantly on a time scale smaller than the time it takes the
reverse shock to cross the ejecta shell (SWL04), and is less
significant when they are slow cooling \citep{MI03}. As mentioned in
footnote \ref{SC}, slow cooling is required in order to have a
detectable radio flare emission, and is therefore the relevant case
for us. Second, the effects of propagation in plasma decrease with
time at a fixed observed frequency $\nu$ \citep{MI03} and are
significantly smaller at the time of the radio flare ($t\sim
1\;$day) compared to the time when the reverse shock is still going
on ($t_{\rm dec}\sim 10^2\;$s). Furthermore, $\Pi_L<1\%$ is expected
at $\nu>\nu_{\rm sa}$ \citep{MI03}, which seems to be the case for
the radio flare emission that we consider in this paper.

Another related effect that might cause depolarization (SWL04) is a
different amount of Faraday rotation for different frequencies at a
given point on the image (which would be hard to resolve because of
the finite instrumental spectral resolution) or at the same observed
frequency between different points on the image (because of the
different path lengths through the emitting plasma and different
Doppler factor corresponding to different frequencies in the local
rest frame of the emitting plasma). The $\Delta\nu/\nu$ which gives
a Faraday rotation of $2\pi$ for a cold plasma is
$|\Delta\nu/\nu|=\pi^2 m_e^2 c^2 (\nu')^2/(e^3n_eB'W')$ where $W'$
is the width of the emitting region in the local frame (SWL04). For
a relativistic plasma with a typical electron random Lorentz factor
of $\gamma_e$ this expression should be multiplied by
$\sim(\ln\gamma_e)/\gamma_e^2$ (Matsumiya \& Ioka 2003; SLW04) so
that Faraday rotation and the resulting depolarization are
suppressed. SWL04 consider fiducial parameters that result in fast
cooling, and obtain that a significant depolarization is possible up
to observed frequencies of $\nu\sim 10^{15}\;$Hz at $t\lesssim
t_{\rm dec}$. Since we consider $\nu\sim 10^{10}\;$Hz and Faraday
rotation scales as $\nu^{-2}$, we need to suppress Faraday rotation
by some 10 orders of magnitude compared to their estimate.

Since a detectable radio flare requires slow cooling of the reverse
shock, this would imply a smaller density and magnetic field in the
local frame, which would reduce the Faraday rotation at $t\lesssim
t_{\rm dec}$. Furthermore, SLW04 used $t_{\rm dec}=10\;$s while the
radio flare occurs at $t\sim 10^5\;$s. We have $n_eW'\propto R^{-2}$
and for the BM solution $W'\propto R/\gamma\propto R^{9/2}$ and
$B'\propto 1/RW'\propto R^{-11/2}$ so that $n_eB'W'\propto
R^{-15/2}\propto t^{-15/16}$, which reduces the Faraday rotation by
$\sim 4$ orders of magnitude between $t_{\rm dec}$ and the time of
the radio flare. Also, at the time of the radio flare $\gamma\sim
5-10$ compared to $\sim 10^{2.5}$ at $t_{\rm dec}$ so that the same
observed frequency corresponds to a higher frequency in the local
rest frame of the emitting plasma ($\nu\approx\gamma\nu'$). This
reduces Faraday rotation by $\sim 3-4$ orders of magnitude. Finally,
the reverse shock electrons must still be relativistic during the
radio flare in order to emit the observed synchrotron radiation, and
this will suppress the Faraday rotation by a factor of
$\sim(\ln\gamma_e)/\gamma_e^2$ (SWL04). Altogether, it seems that
over a large part of the relevant parameter space there will not be
significant depolarization of the radio flare emission due to
different amounts of Faraday rotation at different points on the
image, although it is hard to rule out some degree of depolarization
for certain regions of parameter space.

\section{discussion}
\label{dis}

If the magnetic field is carried out from the source, then at large
radii it would be almost completely in the plane perpendicular to
the radial direction, since the radial component of the magnetic
field scales as $R^{-2}$ while the two transverse components scale
as $R^{-1}$ if the magnetic field is initially tangled on small
scales. If the flow and the magnetic field configuration are axially
symmetric then $B_\phi\propto R^{-1}$ while $B_\theta\propto R^{-2}$
so that $B_\phi$ will dominate at large radii and the magnetic field
will be toroidal. If the ratio of electromagnetic to kinetic energy
is $\sigma\lesssim 1$, then the magnetic field is sub-dominant
dynamically and the plasma can in principle keep it tangled on small
scales within the plane normal to the radial direction, if it is
initially tangled on such small scales, and its exact configuration
is not obvious from theoretical considerations. For example, it
could be ordered on angular scales $\theta_B\ll\theta_j$ where
$\theta_j$ is the half-opening angle of the jet. On the other hand
it might be ordered over the whole jet, possibly in a toroidal
configuration. If, on the other hand, the GRB outflow is Poynting
flux dominated (i.e. $\sigma\gg 1$) then the magnetic field is
dynamically dominant and can move the plasma around and arrange
itself in a toroidal configuration over the whole jet
\citep{LB03,LB04}. Thus an ordered toroidal magnetic field is
expected in the latter type of model.

These are the magnetic field configurations that are expected before
the prompt gamma-ray emission. For $\sigma\lesssim 1$ the gamma-ray
emission is likely due to internal shocks, which may cause a shock
produced magnetic field that is expected to be random within the
plane of the shock, so that it can add a random component to an
initially ordered component. A similar effect can also occur in the
reverse shock. The random component ($B_{\rm rnd}$) can reach values
close to equipartition, so that it might be dominant over the
ordered component ($B_{\rm ord}$) if the latter is well below
equipartition ($\sigma\ll 1$). However, for $\sigma\sim 1$ with an
initially ordered magnetic field, the shock produced random
component can at most be comparable to the ordered component.

In Poynting flux dominated models, where initially $\sigma\gg 1$,
there would be no reverse shock and therefore no radio flare if
$\sigma$ remains $\gg 1$ after the prompt gamma-ray emission. Thus,
in order to get a radio flare we need $\sigma\lesssim 1$. This might
still be possible in such a model if after the dissipation of
electromagnetic energy (magnetic reconnection) that causes the
gamma-ray emission, the value of $\sigma$ decreases to $\sim 1$.
Furthermore, the dissipation of magnetic energy (magnetic
reconnection) that gives rise to the prompt gamma-ray emission in
this type of models can also change the local configuration of the
magnetic field in the emitting region, so that it might not be
perfectly toroidal, and could also have a random component that
could potentially be comparable in strength or possibly even exceed
the ordered component.

A combination of an ordered and a random magnetic field component
could decrease the resulting degree of polarization, compared to
that for a purely ordered magnetic field, by a factor of
$\sim\eta/(1+\eta)$ where $\eta\approx\langle B_{\rm
ord}^2\rangle/\langle B_{\rm rnd}^2\rangle$ \citep{GK03}. Thus we
obtain $\eta\lesssim 0.2$ for GRB 991216, where our upper limit on
the polarization is $\gtrsim 6$ times smaller than the predicted
value for a purely toroidal magnetic field. In other words, a
sufficiently sub-dominant ordered toroidal magnetic field together
with a larger random magnetic field component is required. This
requirement is not trivial when starting from an ordered toroidal
magnetic field close to the equipartition value, since the shock
produced random field component would not exceed equipartition and
thus would typically be at most comparable to the ordered component
($\eta\gtrsim 0.5$).

\section{Conclusions}
\label{conc}

We have derived upper limits on the linear and circular polarization
of the radio flare emission discovered from GRBs to date. Our
results are summarized in Table 1. There is a reasonably good case
that the radio flare emission in GRBs 990123, 991216, and 020405
indeed arises from the original ejecta that was shocked by the
reverse shock and then cooled adiabatically (as discussed in \S
\ref{RS}). This emission is also most likely predominantly
synchrotron radiation. Therefore, our upper limits on the
polarization can be used to constrain the magnetic field structure
in the ejecta of these GRBs.

Propagation effects outside of the emitting region might decrease
the measured polarization below the value of the intrinsic
polarization (see discussion in \S \ref{obs}). We have demonstrated
that the relatively small source size, $\lesssim 10^{17}\;$cm, at
the time of the radio flare makes it very unlikely that gradients in
the rotation measure across the image due to the magnetic field in
the ISM of the host galaxy would cause significant depolarization.
Depolarization due to a possible propagation in a molecular cloud or
in the immediate circumburst medium might cause larger gradients in
the rotation measure across the image, but still require extreme
conditions in order to cause significant depolarization.

Propagation effects in the plasma inside the source might cause a
different amount of Faraday rotation at different points on the
image which may cause depolarization since the image is not
resolved. For a fixed observed frequency, such a depolarization is
much smaller during the radio flare compared to $t\lesssim t_{\rm
dec}$ (see discussion in \S \ref{prop_int}). Since Faraday rotation
scales as $\nu^{-2}$ it might still be important in the radio for
some parameter values, although it should not be very important for
a large part of the relevant parameter space. Thus a more thorough
investigation of this effect and its magnitude over the relevant
parameter space is called for. Keeping this caveat in mind, we
continue and examine the implications of our upper limits on the
polarization under the assumption that there was no significant
depolarization.

We have compared our upper limits to the predictions of different
theoretical models. Models where the magnetic field is produced in
the shock would in most cases produce a polarization below our upper
limits. Furthermore, they are expected to produce a polarization
similar to that of the afterglow emission, if indeed the dominant
magnetic field in the afterglow is shock produced and if the
magnetic field configuration behind the afterglow shock is similar
to that behind the reverse shock. For lines of sight near the edge
of the jet the expected polarization might in some cases exceed our
upper limits, if the magnetic field behind the shock is maximally
anisotropic (i.e. random within the plane of the shock or ordered in
the direction normal to the shock). This might suggest a more
isotropic magnetic field configuration behind the shock if
comparable or better upper limits will be found in a larger sample
of GRBs, similar to the current situation with the afterglow
emission where linear polarization has been measured in several GRBs
and was found to be $\Pi_L\lesssim 3\%$ in all cases, perhaps with
one exception -- GRB 020405 where a sharp spike in the polarization
($P=9.9\pm 1.3\%$ at $t=1.3\;$days) was reported
\citep{Bersier03}.\footnote{This optical polarization spike occurred
at a similar time to the measurement of the radio flare emission
from the same GRB which we used in order to derive the upper limit
on its polarization (see Table 1). This is probably a coincidence.}
We note that an almost simultaneous polarization measurement by a
different group \citep{Masetti03} resulted in a significantly lower
polarization, $P=1.5\pm0.4\%$.

Our upper limits on the polarization put much stronger constraints
on models in which there is an ordered magnetic field in the ejecta.
A magnetic field that is roughly uniform across the whole jet would
produce tens of percent of polarization, which is inconsistent with
our upper limits. If the magnetic field is ordered over patches of
angular scale $\theta_B$, which are mutually incoherent, then our
upper limits on the polarization put upper limits on $\theta_B$. The
tightest constraint is for GRB 9901216, for which we find
$\theta_B\lesssim 10^{-2}\;$rad (see \S \ref{imp}).

The polarization light curves for a toroidal magnetic field in the
ejecta, together with our upper limits, are shown in Fig.
\ref{fig1}. For a uniform jet, our upper limits on the polarization
constrain our viewing angle toward GRB 991216 to be $\theta_{\rm
obs}/\theta_0\lesssim 0.4-0.55$ where this range roughly covers the
uncertainty in the dynamics of the ejecta at $t>t_{\rm dec}$ (as
discussed in \S \ref{imp}). These values are for a conservative
value of the spectral slope, $\alpha=-1/3$ which corresponds to
$P_{\rm max}=1/2$. For a structured jet with a toroidal magnetic
field, $P\sim P_{\rm max}$ is expected at $t\sim t_j$ (which applies
for all the upper limits given in Table 1). Therefore, this model
appears to be inconsistent with our upper limits on the
polarization.

\acknowledgments

We thank Ehud Nakar, Tsvi Piran, Davide Lazzati and Elena Rossi for
useful discussions. This research was supported by US Department of
Energy under contract number DE-AC03-76SF00515 (J.G.). The National
Radio Astronomy Observatory is operated by Associated Universities,
Inc., under cooperative agreement with the National Science
Foundation.

\appendix

\section{Polarization of a Uniform Jet with a Toroidal Magnetic Field}
\label{app1}

Here we derive the polarization of a uniform jet with an ordered
toroidal magnetic field, by generalizing the results of
\citet{Granot03} for a uniform transverse magnetic field. The
emission is assumed to arise from a section of a thin spherical
shell moving radially outward with a bulk Lorentz factor $\gamma\gg
1$, that lies within a cone of half opening angle $\theta_j$, which
represents the jet. For simplicity, the emission is integrated over
the jet at a fixed radius and the the differences in photon arrival
time from different angles $\theta$ from the line of sight are
ignored. An integration over the equal arrival time surface of
photons to the observer might introduce small quantitative
differences, but the results should be qualitatively similar, as we
verified by comparing our results to those of \citet{Lazzati04b}.
The emission in the local rest frame of the shell (where quantities
are denoted by a prime) is taken to be uniform across the jet (hence
a uniform jet), and depends only on the angle $\chi'$ between the
direction of the emitted radiation, $\hat{\bf n}'$, and the local
direction of the magnetic field, $\hat{\bf B}'$.
The polarization position angle at a given point on the jet makes an
angle of $\theta_{p,B}=\phi+\arctan\{[(1-y)/(1+y)]\cot\phi\}$ from
the local direction of the magnetic field, $\hat{\bf B}$
\citep{GK03}, where $y\equiv(\gamma\theta)^2$, and $\phi$ is the
angle between $\hat{\bf B}$ and the direction from the line of sight
to that point on the jet.

The Stokes Parameters are given by $(U,Q)/IP_{\rm max}=\int d\Omega
I_\nu(\sin 2\theta_p,\,\cos 2\theta_p)/\int d\Omega I_\nu$, where
$\theta_p$ is measured from some fixed direction, which for
convenience we choose to be the direction from the jet symmetry axis
to the line of sight. We have $d\Omega\propto d\varphi dy$ where
$\varphi$ is the azimuthal angle around the line of sight measured
from the direction between the jet axis and the line of sight. Let
us use the notations $q\equiv\theta_{\rm obs}/\theta_0$,
$y_j=(\gamma\theta_j)^2$, $y_\pm=(1\pm q)^2 y_j$, and
$a\equiv\theta/\theta_{\rm obs}=q^{-1}(y/y_j)^{1/2}$. We have
$I_\nu=I'_{\nu'}(\nu/\nu')^3$ with
$I'_{\nu'}\propto(\nu')^{-\alpha}(\sin\chi')^\epsilon$,
$\nu/\nu'\approx 2\gamma/(1+y)$ and
\begin{equation}
\sin^2\chi'=1-(\hat{\bf n}'\cdot\hat{\bf B}')^2 \approx
\left(\frac{1-y}{1+y}\right)^2\cos^2\phi+\sin^2\phi \approx
\left(\frac{1-y}{1+y}\right)^2 +
\frac{4y}{(1+y)^2}\frac{(a+\cos\varphi)^2}{(1+a^2+2a\cos\varphi)}\ ,
\end{equation}
where $\hat{\bf n}'$ is the direction in the local frame of a photon
that reaches the observer. We also find that
\begin{equation}
\theta_p=\varphi - \arctan\left[\left(\frac{1-y}{1+y}\right)
\frac{\sin\varphi}{(a+\cos\varphi)}\right]\ .
\end{equation}
From symmetry considerations $U=0$ and $P=|Q|/I$. The direction of
the polarization vector on the plane of the sky is along the line
connecting the jet symmetry axis and the line of sight. The degree
of polarization is given by
\begin{equation}
\frac{P}{P_{\rm
max}}=\left[\Theta(1-q)\int_0^{y_{-}}\frac{dy}{(1+y)^{3+\alpha}}\int_0^{2\pi}d\varphi
(\sin\chi')^\epsilon\cos 2\theta_p +
\int_{y_{-}}^{y_{+}}\frac{dy}{(1+y)^{3+\alpha}}\int_{\Psi_1}^{2\pi-\Psi_1}d\varphi
(\sin\chi')^\epsilon\cos 2\theta_p\right]
\end{equation}
$$
\times\left[\Theta(1-q)\int_0^{y_{-}}\frac{dy}{(1+y)^{3+\alpha}}\int_0^{2\pi}d\varphi
(\sin\chi')^\epsilon +
\int_{y_{-}}^{y_{+}}\frac{dy}{(1+y)^{3+\alpha}}\int_{\Psi_1}^{2\pi-\Psi_1}d\varphi
(\sin\chi')^\epsilon\right]^{-1}\ ,
$$
where $\Theta(x)$ is the Heaviside step function,
$\cos\Psi_1=[(1-q^2)y_j-y]/[2q(y_jy)^{1/2}]$, $P_{\rm
max}=(\alpha+1)/(\alpha+5/3)$, and for producing the results shown
in Fig. \ref{fig1} we use $\epsilon=1+\alpha$ \citep{Granot03}.

In order to produce polarization light curves a simple model of
$\gamma(t)$ is added, where $t$ is the observed time. For a jet with
no lateral spreading going into a uniform density medium
$\gamma\propto t^{-\xi}$ where $\xi=3/8+(g-3/2)/[4(2g+1)]$ and
$\gamma\propto R^{-g}$ (see \S \ref{dyn}). We identify the jet break
in the optical light curve with $\Gamma\theta_j=1$ where $\Gamma$ is
the Lorentz factor just behind the forward shock. Thus
\begin{equation}
y_j^{1/2}=\gamma\theta_j=\left(\frac{t_j}{t_{\rm
dec}}\right)^{[g-(3-k)/2]/[(4-k)(2g+1)]}
\left(\frac{t}{t_j}\right)^{-g/(2g+1)}\
.
\end{equation}
As discussed in \S \ref{dyn} we have $(3-k)/2<g<7/2-k$. For the
lower limit $g=(3-k)/2$ which corresponds to $\gamma=\Gamma$ we have
$y_j=(t/t_j)^{-(3-k)/(4-k)}$. For the upper limit $g=7/2-k$ we have
$y_j=(t_j/t_{\rm dec})^{1/[2(4-k)]}(t/t_j)^{-(7-2k)/[2(4-k)]}$.

\end{document}